\documentclass[prb,preprint,showpacs,superscriptaddress]{revtex4}
\usepackage{graphicx}

\begin{document}
\title{\bf  Structural, electronic and magnetic properties of
Mn-doped GaAs(110)
surface}
\author{A. Stroppa}
\email{alessandro.stroppa@univie.ac.at}
\altaffiliation[Presently at: Institute of Material Physics,
University of Vienna, Sensengasse 8/12, A-1090 Wien, Austria and
Center for Computational Materials Science (CMS), Wien, Austria ]{ }
\affiliation{Dipartimento di Fisica Teorica, Universit\`a di
Trieste,\\ Strada Costiera 11, I-34014 Trieste, Italy}
\affiliation{INFM DEMOCRITOS National Simulation Center, Trieste, Italy}

\date{\today}

\begin{abstract}

First principles total-energy pseudopotential calculations have been
performed to investigate STM images of the (110) cross-sectional
surface of Mn-doped GaAs. We have considered configurations with Mn
in interstitial positions in the uppermost surface layers with Mn
surrounded by As (Int$_{As}$) or Ga (Int$_{Ga}$) atoms. The
introduction of Mn  on the GaAs(110) surface results in a strong
local distortions in the underlying crystal lattice, with variations
of interatomic distances up to 3\% with respect to unrelaxed ones.
In both cases, the surface electronic structure is  half-metallic
(or \emph{nearly} half metallic) and it strongly depends on the
local Mn environment. The nearby Mn atoms show an induced
spin-polarization resulting in  a ferromagnetic Mn--As and
antiferromagnetic Mn--Ga configuration. The simulation of the STM
images show very different pattern of the imaged Mn atom, suggesting
that they could be easily discerned by STM analysis.
\end{abstract}


\maketitle

\newpage
\section{Introduction}

The easy integration of ferromagnetism with semiconducting
properties in the same host material provided by Diluted Magnetic
Semiconductors (DMS's) has been considered an important breakthrough
in the semiconductor microelectronics. This is mainly due to the
 unprecedented opportunity to create a new class of
device which would combine the spin degree of freedom to process, to
transfer as well as to store information. \emph{Spintronics} is the
emergent technology which exploits the quantum propensity of
electrons to spin as well as making use of their charge
state.\cite{Spintr1,Spintr2,Spintr3}

The discovery of ferromagnetism in Mn-doped GaAs semiconductor has
become a milestone in spintronic revolution: Mn$_{x}$Ga$_{1-x}$As
 alloys are directly related to the existing GaAs technology,
resulting in the practical realization of device structures
 combining ferromagnetic and nonmagnetic layers.\cite{reviewOhno}

There are several possibilities for a single Mn to be incorporated
in the GaAs. It can occupy or the cation site (substitutional Mn,
Mn$_{Ga}$) either the anion site (As antisite, Mn$_{As}$); it can
also occupy interstitial sites, as reported by K.M. Yu \emph{et al.}
\cite{interst5} Further, other structural defects could be present
in the alloy, such as As antisite (As$_{Ga}$). The fraction of Mn
dopants occupying one or another location  depend on the growth
conditions and techniques .\cite{defectth}

The Curie temperature (T$_{c}$) is a key parameter in designing
room-temperature spintronic devices. The highest T$_{c}$ reachable
for  Mn$_{x}$Ga$_{1-x}$As up to  few years ago was 110
K,\cite{reviewOhno} i.e. rather low for practical technological
purposes. It has been shown that interstitial Mn atoms have a
crucial role in magnetic properties of the
samples.\cite{annea1,PRLMN} An intense experimental and
theoretical efforts have been pursued in the last years in order to
understand the physics of this material and how to raise the Curie
temperature.

Nowadays, a new method has been proposed as alternative to the
growth by Molecular Beam Epitaxy (MBE) of bulk Mn$_{x}$Ga$_{1-x}$As
random alloy: the dopant atoms are incorporated in the sample in
such a way to give rise to a Dirac's $\delta$ function concentration
profile (with locally high dopant concentration) along the grow
direction ($\delta$-doping).\cite{Delta} Remarkably, an important
enhancement of T$_{c}$ is obtained  in these $\delta$-doped
samples\cite{HighTC} (the highest T$_{c}$
 obtained so far with $\delta$-doped sample is 250 K).\cite{HighTC}
 Very recently,  Mn $\delta$-doped GaAs samples in (001) direction have also been  grown
  at TASC Laboratory in Trieste.\cite{Modesti}

Therefore clarifying  the  site geometry and the local environment
of impurities in $\delta$-doped GaAs:Mn  should shed light on the
understanding and the optimization of the magnetic properties of the
system. From the experimental point of view, this study can be
pursued with cross-sectional Scanning Tunneling Microscopy (XSTM):
the  Mn-doped GaAs samples are  cleaved
 along the natural (110) cleavage plane and then analyzed by STM microscopy.

In recent years, several XSTM studies on Mn$_{x}$Ga$_{1-x}$As alloys
have been performed but the local environment (and preferential
geometric site) of defects has not been clarified
yet.\cite{Mikkelsen,Sullivan,yakunin1,yakunin2} From the theoretical
point of view, the existing simulated XSTM images have mainly
focused on the characterization of substitutional impurities on
uppermost surface layers, while a complete and detailed
investigation of interstitial impurity on uppermost surface layers
is still lacking thus preventing the possibility of a full
interpretation of the new XSTM images acquired.

Therefore, stimulated by the recent growth and following XSTM
analysis of Mn $\delta$-doped GaAs samples at TASC,\cite{Modesti} we
have performed density functional calculations to investigate the
structural, electronic and magnetic properties of a single Mn
dopant, by focusing our attention on the impurity
\emph{interstitial} surface configurations. We have also simulated
the corresponding STM images.

This paper is organized as follows: in the next section we
describe the computational method; in Sect. 3 we present our
results for the structural, electronic and magnetic properties; in
Sect. 4 we discuss our results for the XSTM images; finally, in
Sect. 5 we draw our conclusions.

\section{Computational details}
Our study has been performed within Density Functional Theory (DFT)
framework in the Local Spin Density Approximation (LSDA) for the
exchange-correlation (XC) functional by using state-of-the-art
first-principles pseudopotential self-consistent calculations, as
implemented in the ESPRESSO/PWscf code.\cite{pwscf} We used the
scheme of Ceperley and Adler\cite{PZ} (with the parametrization of
Perdew and Zunger\cite{PZ1}) for XC functional. Mn atom is described
by an ultrasoft (US) pseudopotential (PP)\cite{Vander} while
norm-conserving PPs have been considered for Ga, As and H atoms.

Test calculations have shown that a kinetic energy cutoff for the
wave functions equal to 22 Ry and a 200 Ry cutoff for the charge
density are sufficient to get well converged results. We estimate
the numerical uncertainty  to be
  $\sim$ 0.01 \ \AA \ for relative atomic displacements and
 $\sim$  0.02 $\mu_{B}$ for the magnetic moments.
The relaxed internal atomic positions
 have been obtained by total-energy and atomic-force minimization using the
 Hellmann-Feynman theorem.\cite{forces}

 The surface is modelled with periodically repeated cell containing one Mn
atom; a (110) slab geometry with
 a 4$\times4$ in-plane periodicity has been used. The simulation cells are
made up of 5 atomic layers and a vacuum region equivalent to 8
atomic layers. The bottom layer has been passivated with Hydrogen
atoms in order to simulate semi-infinite bulk
material.\cite{passivation} In the energy minimization only the
three uppermost layers are allowed to relax, while the others are
 considered bulk-like.

Two different configurations have been considered for Mn on the
surface, namely Int$_{As(Ga)}$ with As (Ga) atoms as nearest
neighbor atoms. In each case, the distances between the Mn atom and
its periodic image on the (110) plane are 15.7 \AA \ along the
[1$\bar{1}$0] and 22.2 \AA \ along [001].

 XSTM images are obtained within
 Tersoff-Hamann model,\cite{Ters2} where the constant current STM images are
 simulated from electronic structure calculations by considering
 surfaces of constant integrated local density of states.

\section{Structural, electronic and  magnetic properties}
\renewcommand{\thesubsection}{\arabic{subsection}}
\subsection{Structural properties}
\begin{figure}[!hbp]
\caption{Schematic side  and top view of the clean GaAs(110)
surface.
 Only the three topmost layers (1$^{st}$ layer is the surface layer)
 are shown in the Figure. In this and other figures,
 black spheres are cations (Ga atoms), grey spheres are anions
 (As atoms).}\label{Fig1}\vspace{1cm}
\includegraphics[scale=.8,angle=0]{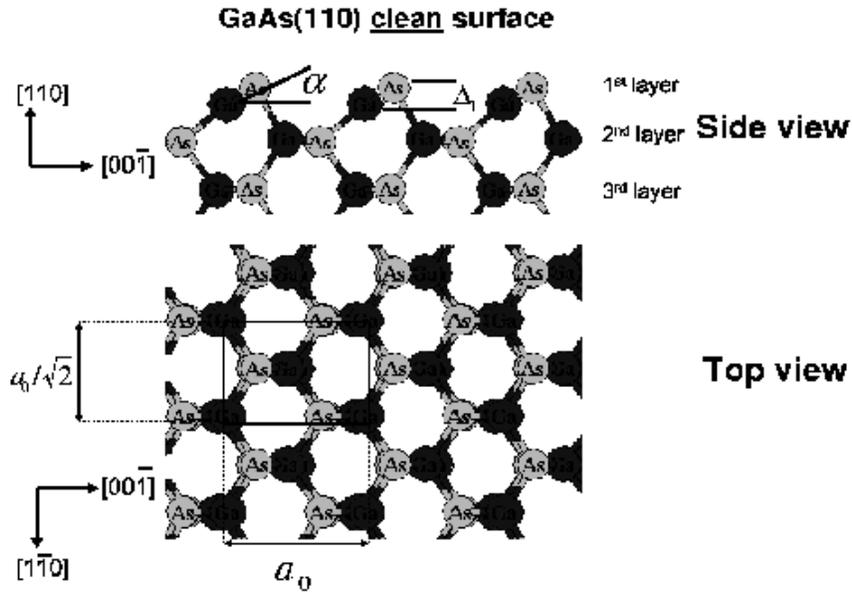}
\end{figure}

The GaAs(110) surface is well known from experimental as well as
theoretical point of view.\cite{report} In fig.\ref{Fig1}, we show a
ball and stick model of the clean surface, side and top views. The
surface unit cell in shown in the top view. In this and the other
figs., black spheres are cations (Ga atoms), grey spheres are anions
(As atoms).

At the top layer, the Ga surface atoms relax inward while the As
atoms are shifted above the surface. Due to overbinding in the LDA
approximation, our theoretical GaAs lattice constant (5.55 \AA) is
smaller than the experimental one (5.65 \AA) but the relevant
calculated structural parameters for the clean surface
 such as $\Delta_{1,\bot}$ (relative displacement of the anion and
 cation positions in the uppermost layer, normal to the surface) and $\alpha$ (the
buckling angle), shown in fig.\ref{Fig1},
  are 0.68 {\AA} and 30.36$^{\circ}$ respectively, which well compare with the
experimental values 0.65$\pm$0.03  \AA\ and
27.4$^{\circ}$\cite{report,exp1} and other theoretical
works.\cite{Miotto,oth1,oth2}

\begin{figure}[!hbp]
\caption{Conventional bulk unit cells representing Mn atom in
tetrahedral-interstitial configurations, surrounded by As atoms
(grey spheres) as nearest neighbors (top part, to the left) and by
Ga atoms (black spheres) as nearest neighbors (top part, to the
right). Bottom part: hexagonal interstitial position with Mn
surrounded by 3 As and 3 Ga as nearest
neighbors.}\label{Fig2}\vspace{1cm}
\includegraphics[scale=.8,angle=0]{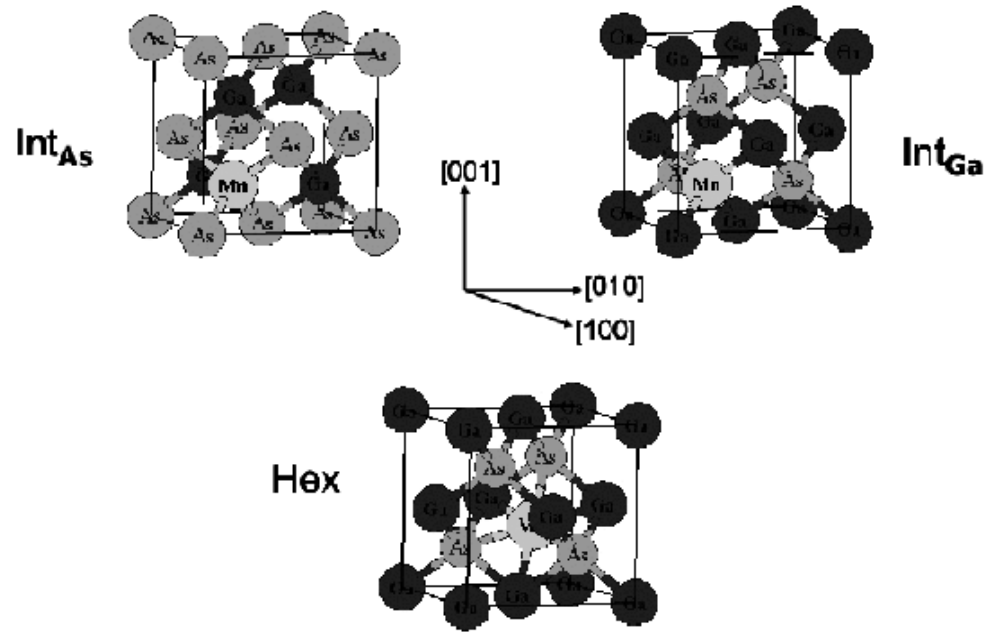}
\end{figure}

In zinc-blende  bulk crystal there are two inequivalent tetrahedral
interstitial positions for Mn which differ in their local
environment: we denote them as Int$_{As}$ or Int$_{Ga}$ according
whether Mn  is surrounded by As or Ga atoms respectively. There is
also an hexagonal interstitial position where Mn is surrounded by
\emph{both} As and Ga atoms. In fig.\ref{Fig2} we show the different
cases.  The tetrahedral interstitial site in the ideal geometry has
four nearest-neighbor (NN) atoms at a distance equal to the ideal
host bond length $d_{1}$ and six next-nearest-neighbor (NNN) atoms
at the distance $d_{2}=\frac{2}{\sqrt 3}d_{1}$, which are Ga(As)
atoms for Int$_{As(Ga)}$, respectively. In the hexagonal
interstitial position the Mn is surrounded by 3 As and 3 Ga atoms at
distance $\sqrt{\frac{11}{12}}d_{1}$. Throughout this work we have
considered only \emph{tetrahedral} interstitial position (the total
energy corresponding to the \emph{hexagonal} interstitial site is
higher by more than 0.5 eV).\cite{Maka1,Maka2,PRLMN,condmat}

\begin{figure}[!hbp]
\caption{Schematic side and top view of the relaxed Int$_{As}$ and
Int$_{Ga}$ configurations. Mn is explicitly
shown.}\label{Fig3}\vspace{1cm}
\includegraphics[scale=.8,angle=0]{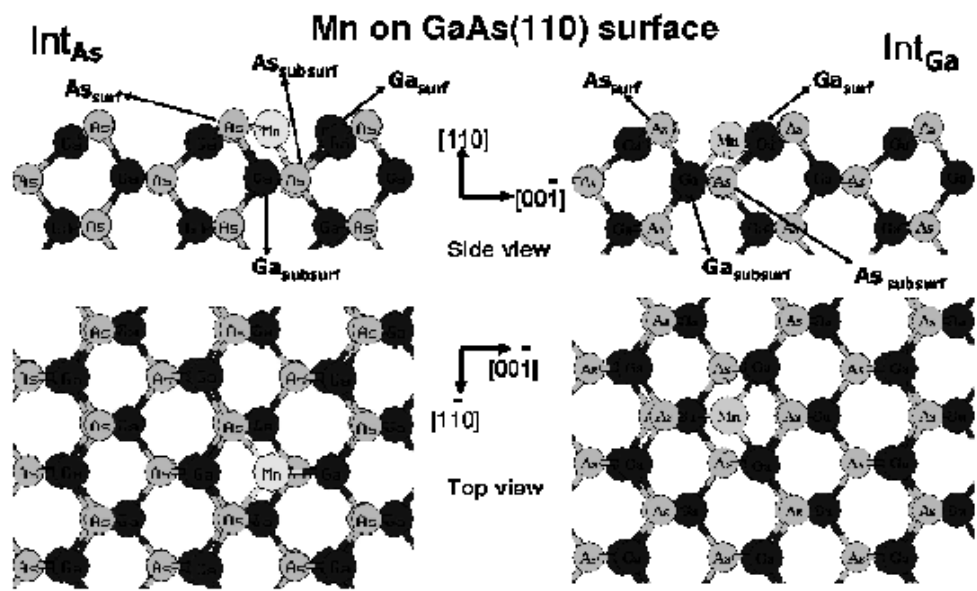}
\end{figure}

In fig.~\ref{Fig3} we show a ball and stick side (a) and top (b)
view of the relaxed Int$_{As}$, Int$_{Ga}$ configurations. Only the
three topmost layers and the atoms closest to Mn are shown. Black
spheres are cations (Ga atoms), grey spheres are anions (As atoms);
Mn is explicitly indicated. In the relaxed structure,
 due to symmetry breaking because of the surface and
 the consequent buckling of the outermost surface layers,
 the NN and  NNN bond lengths are
no longer equal. Furthermore, some relaxed NNs bond lengths turn out
to be longer than NNNs ones. In the following, we do not longer
distinguish among NN and NNN (they are referred simply as NN atoms)
but we simply refer to \emph{surface} and \emph{subsurface} atoms,
as shown in the figure.

The two relaxed configurations differ in energy  by $\sim$ 130
meV/Mn atom (Int$_{Ga}$ is favoured). This is in contrast to the
bulk case, where it has been found that they differ only by $\sim 5$
meV/Mn\cite{bulkint} and Int$_{As}$ instead is slightly favored. We
have tested the reliability of our final relaxed interstitial
configurations by considering different starting geometries (details
in Ref.~\onlinecite{InpressStroppa}), other than the simple ideal
(110) truncated bulk. In all cases, the final relaxed configuration
is the same.

The atoms with the most sizeable displacements from the ideal zinc
blende positions are the Mn impurities and their neighbors, on
surface or subsurface.
 In Tab.~\ref{tab1} we report the inward/outward
 relaxations  respect to the ideal (110) surface plane.

In Int$_{As}$, Mn relaxes outward by $\sim$ 0.06 \AA \ and
As$_{surf}$ (As$_{subsurf}$) move upwards (downwards). On the other
hand, the Ga atoms (both on surface and subsurface) are shifted
towards the bulk.

In Int$_{Ga}$, Mn relaxes inward by $\sim$ 0.32 \AA; the Ga$_{surf}$
and Ga$_{subsurf}$ atoms are displaced downwards while the
As$_{surf}$ (As$_{subsurf}$) atom moves upwards (downwards). In
summary, both in Int$_{Ga}$ and Int$_{As}$, cations (surface and
subsurface) close to Mn move downwards, while anions upwards or
downwards according whether they are on surface or subsurface. The
net result is a local reduction of the surface buckling with respect
to the clean unperturbated surface, more than 30 \% and 40 \% for
Int$_{As}$ and Int$_{Ga}$ respectively, with a net local buckling of
about 0.46 \AA \ for Int$_{As}$ and 0.40 \AA \ for Int$_{Ga}$.
As far as the interatomic distances  between Mn and the nearest
atoms are concerned (Tab.~\ref{tab1}), they are in general longer
than the ideal bulk value by $\sim$ 2-3 \%; the distances between Mn
and more distant atoms are shorter than the bulk cases, except for
Ga$_{subsurf}$ in Int$_{As}$, as it can be seen in Tab.~\ref{tab1}.

\subsection{Electronic properties}
\begin{figure}[!hbp]
\caption{Density of States (DOS) projected into surface (continuous
line) layer for Int$_{As}$ (to the left) and Int$_{Ga}$ (to the
right). Dashed line corresponds to the DOS for the clean surface. Mn
projected DOS is also shown (grey filled area). The Fermi level is
set to zero eV.}\label{Fig4}\vspace{1cm}
\includegraphics[scale=.8,angle=0]{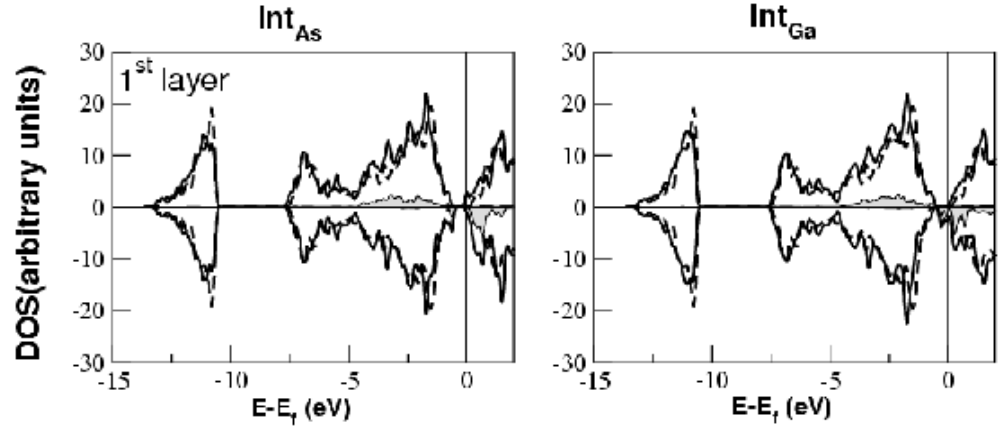}
\end{figure}

In fig.~\ref{Fig4}, we show the Density of States projected onto
surface layer (PDOS); the continuous lines refer to Int$_{As}$ or
Int$_{Ga}$ while the dashed lines refer to the clean GaAs (110)
surface. DOS for Int$_{As}$(Int$_{Ga}$) are shown to the left
(right) side; the Fermi level (E$_{f}$) is  set to zero eV. The $d$
Mn projected DOS is also shown (grey area). The positive and
negative DOS correspond to spin-up and spin-down components. First
of all, in both Int$_{As}$ and Int$_{Ga}$, the DOS curves for
Int$_{As}$ and Int$_{Ga}$ are very close to those corresponding to
the clean surface case, but they differ in the energy region  around
E$_{f}$. An energy gap around E$_{f}$ is present in both majority
and minority DOS. In Int$_{As}$ the majority and minority spin gaps
overlap and almost coincide, maintaining the surface semiconducting
with a gap of about $\sim$ 0.2 eV. In Int$_{Ga}$, instead, majority
and minority spin gaps are quite different:  $\sim$ 0.3 eV for
 the majority component and $\sim$ 0.1 eV
 for the minority component. The perturbation is weak on the valence
  band and stronger on the conduction band. The main
  difference between Int$_{As}$ and Int$_{Ga}$ DOS curve
  concerns  a peak in the minority component in Int$_{Ga}$ around the Fermi energy
  (in Int$_{As}$ it is shifted by 0.3-0.4 eV below the Fermi energy)
 which  reduces the gap in Int$_{Ga}$.

 In both systems, the Fermi level lies in the lower tail of the
 conduction band thus indicating that interstitial Mn impurity behaves as
 a donor, like in the bulk case.\cite{Maka2}

At variance with  the bulk case, where the calculated DOS for the
two tetrahedral interstitial positions are almost the
same,\cite{Maka2} thus indicating a week influence of the nearest
neighbors on the interstitial Mn in the two configurations, the
difference between surface
 Int$_{Ga}$ and  Int$_{As}$ cases is more sizeable,
 indicating a stronger effect of the local environment.

The PDOS almost recover the bulk features already in the second
layer (not shown in fig.~\ref{Fig4}). Therefore, the introduction of
Mn results  in a perturbation of the electronic properties mostly
localized on the first layer and  strongly depending on the local
environment.

As far as  the $d$ states are concerned, we observe that their
contribution to the occupied majority spin component is by far
larger than their contribution to the minority spin. However, their
overall weight in the GaMnAs system is negligible and the valence
band is in practise almost non spin-polarized (as observed above).
In both cases, the Mn spin-up $d$ states are occupied and quite
similar in shape while the spin-down $d$ states are almost
unoccupied and they have a different shape, especially around the
Fermi level.

In conclusion, the two Mn local environment give rise to a quite
different surface electronic structure, with the differences mainly
localized around the Fermi level.

\subsection{Magnetic properties}\label{magnetism}

In the following, we analyze the magnetic properties.
 The total and absolute magnetization in the supercell are
  different in the two configurations. They are equal to 4.23
  and 4.84 $\mu_{B}$ in Int$_{As}$ and to 3.41 and
  4.71 $\mu_{B}$ in Int$_{Ga}$.
 The difference between total and absolute magnetization
 corresponds to the presence of region of negative
  spin-density in the unit cell; this difference is higher in Int$_{Ga}$ than in
  Int$_{As}$, suggesting higher (absolute)
   values and/or more extended region
  of negative spin-density in the former than in the latter.
  It also
  justifies the smaller total magnetization of Int$_{Ga}$
  with respect to Int$_{As}$. This is a clear evidence that the induced magnetization is
  strongly influenced  by the local Mn environment.

  Interesting information can be gained by looking at the individual
  atomic magnetic moments obtained as the difference
  between the calculated majority and minority Lowedin
  charges.\cite{Lowedin}
  The results have been reported elsewhere.\cite{stroppaperessi}
  The highest value of Mn spin-polarization is found in Int$_{As}$ (3.96
 $\mu_{B}$)  while it is slightly lower in Int$_{Ga}$ (3.67
 $\mu_{B}$). The Mn magnetic in Int$_{As}$ is almost integer in agrement with
 the existence of a clear gap in the Mn-projected DOS and the
 unoccupied states just cutting the Fermi energy.
 It is worth noting that our calculated Mn magnetic moments
  are larger than those corresponding to the Interstitial Mn in the bulk
  and they are rather close to the value indicated for ferromagnetically
  coupled substitutional Mn impurities on the Ga sublattice in bulk
  GaAs.
   In fact, ab--initio calculations\cite{Maka1,Sanyal,Wu}
   report a Mn magnetic moment
   for \emph{bulk} Int$_{As}$
   equal to 2.70 $\mu_{B}$.
   A recent experimental work\cite{Gambardella} show that Mn impurities on GaAs(110)
    surfaces have magnetic moments significantly larger compared
  to the bulk case.  The experimental and theoretical results
   would suggest in general an enhancement
   of the Mn magnetic moments due to
  surface effects. Our calculations, compared with previous bulk DFT
  studies,\cite{Maka1,Sanyal}
  support this indications.

For Int$_{As}$, the As$_{surf}$ and As$_{subsurf}$ atoms have a
ferromagnetic coupling to Mn, with a small magnetic moment equal to
0.05 $\mu_{B}$. The induced polarization in more distant As atoms is
totally negligible. The Ga$_{surf}$ atoms couple
antiferromagnetically  with Mn with an induced polarization on it
equal to  -0.14 $\mu_{B}$. Other atomic moments are negligible.

As far as the Int$_{Ga}$ configuration is concerned, a negative
magnetic moment is induced on  Ga$_{surf}$ (-0.17 $\mu_{B}$) while
the Ga$_{subsurf}$ atoms have a negligible polarization. The
As$_{surf}$ shows only a negligible polarization, while it is
positive and equal to 0.05 $\mu_{B}$ for As$_{subsurf}$.

Our results for the magnetic properties can be summarized as
follows: in both cases, the cations couple antiferromagnetically to
Mn spin moment while anions couple ferromagnetically. Furthermore,
only surface cations are spin polarized, while both surface and
subsurface anions do polarize.

\section{STM Images}
\begin{figure}[!hbp]

\caption{Simulated STM images of isolated Mn interstitial in
GaAs(110) surface, with As NNs (to the left) and Ga NNs (to the
right). Top panels: ball-and-stick model of the relaxed surface, top
and side view (Ga: black spheres, As: grey spheres). Bottom panel:
simulated STM images for positive bias voltage. The intersection of
the dotted lines locates the position of Mn (projected on the (110)
plane).}\label{Fig5}\vspace{1cm}
\includegraphics[scale=1.0,angle=0]{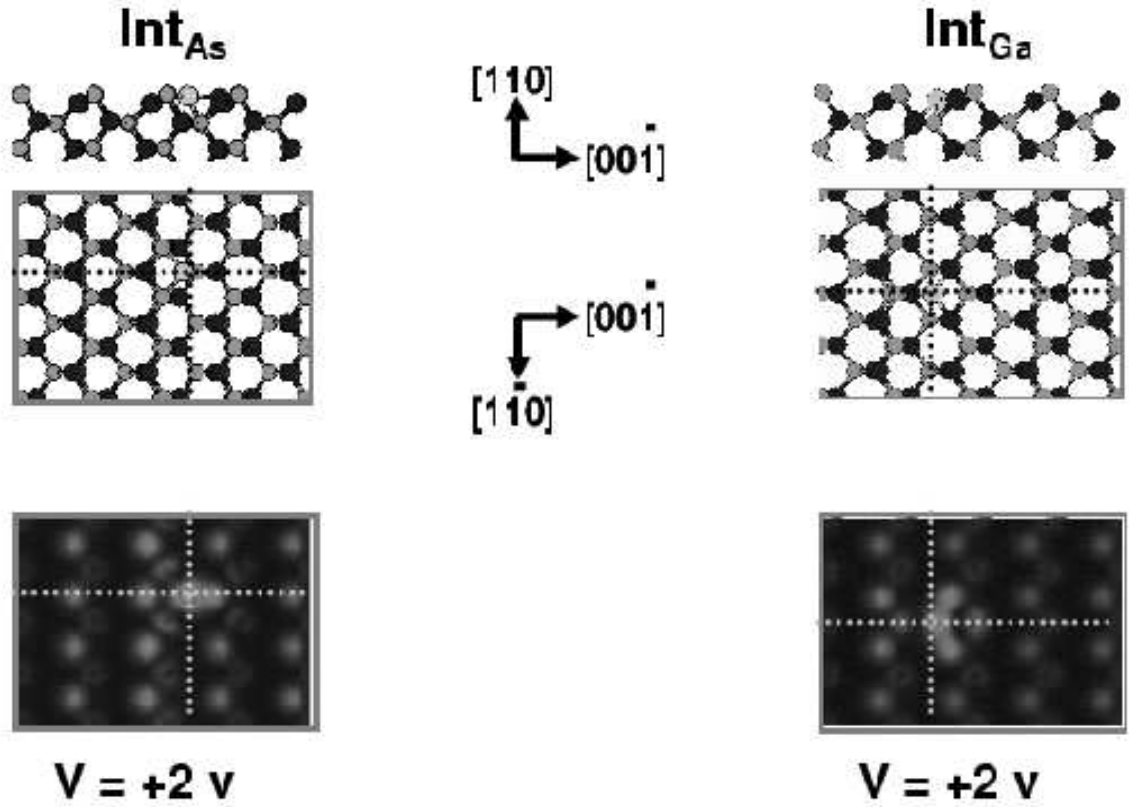}

\end{figure}

In fig.\ref{Fig5} we show the schematic front and side views  of the
relaxed underlying structure lattice and the XSTM images, for empty
states at a reference positive bias voltages ($+$2.0 V). In
Int$_{Ga}$, the two NN surface Ga atoms of Mn appear very bright
with features extending towards the Mn and the atoms in the
neighbourhood also look brighter than normal. For Int$_{As}$, a very
bright elongated spot in the center of the surface unit cell
delimited by As is visible. We would like to point out that the
simulated XSTM images have clearly different shape for the two
geometric configurations, so the two different local coordination
should be distinguished by STM analysis. Further, the simulated STM
images for Int$_{As}$ case well compare  with experimental XSTM
images of the $\delta$-doped samples.\cite{Modesti}

\section{Conclusion}
In summary, we have used first-principles simulations to
characterize Mn interstitial impurity on the GaAs (110) surface.
Strong local distortion on the (110) GaAs surface are introduced by
Mn, especially when it is surrounded by Ga atoms.  In both case, Mn
polarizes the NN and NNN atoms, giving rise to a ferromagnetic
Mn--As and to an antiferromagnetic Mn--Ga configuration. The
simulated STM images show very different shape of the imaged Mn
atom, suggesting that two configuration can be clearly
differentiated by STM analysis. Finally, recent experimental STM
images are qualitatively similar to our simulated one for Int$_{As}$
configuration, suggesting the possible identification of Mn
interstitials surrounded by As atoms in the experimental
samples.\cite{Modesti}

\section{Acknowledgments}
The author would like to thank: S. Modesti, D. Furlanetto and X.
Duan for fruitful discussions; A. Debernardi for providing me his
pseudopotential for Manganese; Computational resources have been
obtained partly within the ``Iniziativa Trasversale di Calcolo
Parallelo'' of the Italian {\em Istituto Nazionale per la Fisica
della Materia} (INFM) and partly within the agreement between the
University of Trieste and the Consorzio Interuniversitario CINECA
(Italy). All the ball-and-stick figures presented here have been
generated by using the Xcrysden Package\cite{Xcrysden}.

{}

\newpage

\begin{table}[!hbp]
\caption{Vertical atomic displacements with respect to ideal zinc
blende bulk positions (first row) and  nearest-neighbor surface
and subsurface relaxed interatomic distances (second row) for Int$_{As}$
(upper part) and Int$_{Ga}$ (lower part); $+/-$ refer to an
downward/upward relaxation; the numbers in round brackets refer to
\emph{unrelaxed} interatomic distances. Units are in \AA.}\label{tab1}
\vspace{1.cm}
\begin{tabular}{||c|c|c|c||}

\hline\hline

\hline

\multicolumn{4}{|c|}{Nearest Neighbor bond-lengths (\AA)}\\
\hline
\multicolumn{4}{||c||}{Int$_{As}$}\\
\hline

As$_{surf}$&As$_{subsurf}$&Ga$_{surf}$&Ga$_{subsurf}$\\

\hline
+0.15 &-0.19 & -0.06&-0.06\\
\hline
2.52(2.40)&2.44(2.40)&2.49(2.78)&2.90(2.78)\\
\hline\hline
\multicolumn{4}{||c||}{Int$_{Ga}$}\\
\hline Ga$_{surf}$ &Ga$_{subsurf}$ &As$_{surf}$
&As$_{subsurf}$ \\
\hline
\hline

-0.22 &-0.24&+0.06&-0.10\\
\hline
2.48(2.40) &2.56(2.40)&2.68(2.78)&2.63(2.78)\\
\hline\hline
\end{tabular}
\end{table}

\clearpage

\clearpage

\clearpage

\clearpage

\clearpage

\clearpage

\vspace{1cm}

\clearpage

%



\begin{thebibliography}{99}
\bibitem{forces} For the optimation of atomic positions we require
 Hellmann-Feynman forces smaller then 0.02 eV\AA$^{-1}$.
\end{thebibliography}
\end{document}